\documentclass[utf8]{frontiersFPHY} 

\setcitestyle{square} 
\usepackage{url,hyperref,lineno,microtype,subcaption}
\usepackage[onehalfspacing]{setspace}

\def\keyFont{\fontsize{8}{11}\helveticabold }
\def\firstAuthorLast{Andreas Ekstr\"om} 
\def\Authors{Andreas Ekstr\"om}
\def\Address{Department of Physics, Chalmers University of Technology, SE-412 96 Gothenburg, Sweden}
\def\corrAuthor{Andreas Ekstr\"om}

\def\corrEmail{andreas.ekstrom@chalmers.se}

\begin{document}
\onecolumn
\firstpage{1}

\title[\textit{Ab initio} models of atomic nuclei: challenges and new
  ideas]{\textit{Ab initio} models of atomic nuclei:\\challenges and
  new ideas}

\author[\firstAuthorLast ]{\Authors} 
\address{} 
\correspondance{} 

\extraAuth{}
\maketitle

\begin{abstract}
This review presents some of the challenges in constructing models of
atomic nuclei starting from theoretical descriptions of the strong
interaction between nucleons. The focus is on statistical computing
and methods for analyzing the link between bulk properties of atomic
nuclei, such as radii and binding energies, and the underlying
microscopic description of the nuclear interaction. The importance of
careful model calibration and uncertainty quantification of
theoretical predictions is highlighted.
   
\keyFont{ \section{Keywords:} Nuclear interactions, chiral effective
  field theory, model calibration, uncertainty quantification,
  optimization, Bayesian parameter estimation.}
\end{abstract}

\section{Introduction}
The \textit{ab initio} approach to describe atomic nuclei and nuclear
matter is grounded in a theoretical description of the interaction
between the constituent protons and neutrons. The long-term goal with
this course of action is to construct models to describe and analyze
the properties of nuclear systems with maximum predictive power. It is
of course well known that the elementary particles of the strongly
interacting sector of the Standard Model are quarks and gluons, not
protons and neutrons. However, since the relevant momentum scales of
typical nuclear structure phenomena are low enough to not resolve the
internal degrees of freedoms of nucleons, it is reasonable to model
the nucleus as a collection of strongly interacting and point-like
nucleons. This idea has inspired significant efforts aimed at
developing algorithms and mathematical approaches for solving the
many-nucleon Schr\"doinger equation in a bottom-up fashion and with as
few uncontrolled approximations as possible, see
e.g. Refs.~\cite{dickhoff2004,lee2009,bogner2010,barrett2013,Carbone:2013eqa,hagen2014,hergert2014},
as well as a multitude of theoretical descriptions of the interaction
between nucleons, at various levels of phenomenology, see
e.g. Refs.~\cite{wiringa1995,machleidt2001,entem2003}, and
Refs.~\cite{vankolck1999, epelbaum2009, machleidt2011} for
comprehensive reviews on (chiral) effective field theory (EFT)
methods. Reference~\cite{Machleidt1989} also offers a historical
account of various approaches to understand the nuclear interaction.

Currently, \textit{ab initio} modelling of atomic nuclei face two main
challenges:
\begin{itemize}
\item We have limited knowledge about the details of the interaction
  between nucleons, which in turn limits our ability to predict
  nuclear properties.
\item Given a microscopic description of the interaction between
  nucleons inside a nucleus, a quantum-mechanical solution of the
  nuclear many-body problem is exacerbated by the curse of
  dimensionality.
\end{itemize}

There is however continuous progress on both frontiers, and attempts
at quantifying the uncertainty of model predictions are beginning to
emerge in the community. Rapid algorithmic advances in combination
with a dramatic increase in available computational resources make it
possible to employ several complementary mathematical methods for
solving the nuclear Schr\"odinger equation. We can nowadays generate
numerical representations of microscopic many-nucleon wavefunctions,
for selected medium-mass and heavy-mass nuclei, with a rather
impressive precision. Although several observables remain beyond the
reach of state-of-the-art models, e.g. most properties associated with
highly collective states, we can still describe certain classes of
observables rather well, such as total ground-state binding energies
and radii, and sometimes low-energy excitation spectra. We are thus
capable of analyzing experimentally relevant nuclei directly in terms
of a quantum mechanical description of the interaction between its
constituent nucleons. Indeed, the list of, sometimes glaring,
discrepancies between theory and experiment furnish some of the most
interesting nuclear physics questions at the moment, see
e.g. Refs.~\cite{elhatisari2015,rosenbusch2015,lapoux2016,lee2019,taniuchi2019,Gysbers2019,soma2019}.
Many of these efforts are aimed understanding the nuclear binding
mechanism, the location of the neutron dripline, the existence of
shell-closures and magic numbers in exotic systems, and the emergence
of nuclear saturation.

State-of-the-art theoretical analyses of experimental data indicate a
large and non-negligible systematic uncertainty in the description of
bulk nuclear observables, see e.g. Ref.~\cite{BINDER2014119}. Given
the high-precision of modern many-body methods, much of this
uncertainty can be traced to the description of the interaction
potential. Although there exists \textit{ab initio} models that
describe nuclei rather well, albeit in a limited domain, it is less
clear why other models sometimes fail. Indeed, the NNLO$_{\rm sat}$
interaction potential~\cite{ekstrom2015} reproduces several key
experimental binding energies and charge radii for nuclei up to mass
$A \sim 50$~\cite{hagen2015,garciaruiz2016,duguet2017,soma2019}, while
the so-called 1.8/2.0 (EM) interaction
potential~\cite{nogga2004,hebeler2011} reproduces binding energies and
low-energy spectra up to mass $A \sim
100$~\cite{hagen2015,hagen2016b,morris2017,simonis2017,liu2019,holt2019}
while radii are underestimated. The origin of the differences between
these potentials is unknown. It is of course the role of nuclear
theory to close the gap between theory and experiment by developing
and refining the theoretical underpinnings of the model. But given the
complex nature of atomic nuclei, there is significant value in trying
to quantify, or estimate, the detailed structure of the observed
theoretical uncertainty. This might provide important clues about
where we should focus our efforts. There exists well-defined
statistical inference methods that can provide additional guidance,
and several ongoing projects are currently focused on applying
statistical computing methods in the field of \textit{ab initio}
modelling. The topic of uncertainty quantification in nuclear physics
has been discussed at a series of workshops on Information and
Statistics in Nuclear Experiment and Theory (ISNET). Recent
developments in this field are documented in the associated focus
issue published in Journal of Physics G~\cite{Ireland_2015}. A second
focus issue has just been announced, and the first few papers are
already published.

In Sec.~\ref{sec:modelling} and Sec.~\ref{sec:model_calibration} of
this paper I will review a selection of recent results and often
applied methods for calibrating \textit{ab initio} models. In
Sec.~\ref{sec:bayes} and Sec.~\ref{sec:future} I will discuss some of
the recently emerging strategies for making progress using statistical
computing and Bayesian inference methods. The aim is to provide an
overview of selected accomplishments in the field of statistical
inference and statistical computing with \textit{ab initio} models of
atomic nuclei. Hopefully, this paper can serve as a brief introduction
to practitioners who wish to learn about ongoing developments and
possible future directions.

As a final remark, in this paper I will try to consistently use the
word \textit{model} when referring to \textit{any} current method for
theoretically describing the properties of atomic nuclei, including
descriptions that claim to be building on more fundamental
underpinnings, such as EFT. One can certainly make a finer distinction
between models, EFTs, and theories. As outlined in
Ref.~\cite{HARTMANN2001267}; theories provide a unified framework,
categorization, and the joint language used for discussions; EFTs
capture physics at a given momentum scale; and models can be used to
study aspects of a theory, increase understanding, and provide
intuition.

\section{Ab initio models of nuclear many-body systems}
\label{sec:modelling}
An \textit{ab initio} model is here defined as a description that is
based on wavefunctions $|\Psi \rangle$ that solve the many-nucleon
Schr\"odinger equation
\begin{equation}
  [\hat{T} + \hat{V}(\vec{\alpha})] |\Psi\rangle = E|\Psi\rangle.
  \label{eq:schrodinger}
\end{equation}
In this schematic representation, $\hat{T}$ is the total kinetic
energy operator for the $A$-nucleon system, $\hat{V}(\vec{\alpha})$ is
the potential energy operator for the interaction between the
nucleons, and $E$ is the total energy of the system in the state
represented by $|\Psi\rangle$. The potential operator term depends on
a set of parameters $\vec{\alpha}$ that governs the strengths of the
various interaction pieces in the potential. Below I will also refer
to these parameters as low-energy constants (LECs). Given a particular
expression for the potential $\hat{V}$, with numerical values for the
parameter vector $\vec{\alpha}$, and a mathematical method to solve
Eq.~\ref{eq:schrodinger} for e.g. the state $|\Psi\rangle$ with lowest
energy, it is in principle possible to quantitatively compute the
expectation value for any observable $\hat{\mathcal{O}}$ with respect
to this state, e.g. its charge radius. Of course the trustworthiness
of the result and its level of agreement with experimental data can
vary dramatically between different models, i.e. combinations of
potentials and many-body methods.

I will denote an \textit{ab initio} model with
$M(\vec{\alpha},\vec{x})$. It is defined as the combination of a
definite expression for the potential $\hat{V}(\vec{\alpha})$, and a
method for solving the Schr\"odinger equation. The vector $\vec{x}$ is
a set of control inputs that specify all necessary settings such as
nucleon numbers, which observable to compute, values of the
fundamental physical constants, and algorithmic settings for the
mathematical method used for solving Eq.~\ref{eq:schrodinger}. Once a
set of numerical values for $\vec{\alpha}$ has been determined, a
subset of the control inputs $\vec{x}$ of the model can be varied to
make model predictions, preferably at some physical setting, for
e.g. exotic nuclei where we cannot easily make measurements. Provided
that the form of the potential operator $\hat{V}$ and relevant
physical constants remain the same, and the model parameters
$\vec{\alpha}$ were calibrated carefully, it is of course possible to
transfer the vector $\vec{\alpha}$ between \textit{ab initio} models
based on different methods for solving the many-nucleon Schr\"odinger
equation. This is also in line with a physical interpretation of the
parameters $\vec{\alpha}$ that elevate them to a status beyond being
simple tunable parameters inherent to a specific model with the sole
purpose of achieving a good fit to calibration data. This will be
discussed further in Sec.~\ref{sec:model_calibration}.

One of the most exciting developments in nuclear theory is that we
nowadays have access to a range of methods for solving
Eq.~\ref{eq:schrodinger} with very high numerical precision for
selected isotopes and observables. This gives us the opportunity to
compare model predictions with experimental data to learn more about
the elusive structure of the interaction between nucleons. However,
such an analyses require careful statistical interpretation of the
theoretical results. In particular a sensible estimate of the
uncertainty associated with a theoretical prediction. Indeed, only
with reliable theoretical errors is it possible to infer the
significance of a disagreement between experiment and theory, which in
turn may hint at new physics.

\subsection{Chiral potentials and the strong interaction between nucleons}
On a fundamental level, the atomic nucleus is a quantum mechanical and
self-bound system of interacting nucleons. In turn these particles are
composed of three quarks whose mutual interactions are described well
by the Standard Model of particle physics. As such, starting from the
Standard Model it should be possible to account for all observed
phenomena also in atomic nuclei, besides possible signals of beyond
Standard Model physics. However, to theoretically understand the
emergence of nuclei from the Standard Model is an open problem, and
linking the quantitative predictions of atomic nuclei to the dynamics
of quarks and gluons is a central challenge in low-energy nuclear
theory. Although, viewing the atomic nucleus as a (color-singlet)
composite multi-quark system is not the most economical
choice. Indeed, the strong interaction, which is the most important
component for nuclear binding and well-described by quantum
chromodynamics (QCD), is non-perturbative in the low-energy region
inhabited by atomic nuclei. Non-perturbative Monte Carlo sampling of
the quantum fields of QCD amounts to a computational problem of
tremendous proportions. This strategy, referred to as lattice QCD, is
expected to require at least exascale resources for a realistic
analysis of even the lightest multi-nucleon systems. Without any
unforeseen disruptive technology, this approach will not provide an
operational method for routine analyses of nuclei. For the cases where
numerically converged results can be obtained, lattice QCD offers a
unique computational laboratory for theoretical studies of QCD in a
low-energy setting, see e.g. Ref.~\cite{Barnea2015,Chang}.

The description of nuclei should nevertheless build on QCD, or the
Standard Model in general. A turning point in the development of
QCD-based descriptions of the nuclear interaction came when EFTs of
QCD~\cite{WEINBERG1979327} arrived also to many-nucleon
physics~\cite{weinberg1990}. An EFT formulates the dynamics between
low-energy degrees of freedom, e.g. nucleons and pions, in harmony
with some assumed symmetries of an underlying theory, e.g. QCD, and
any high-energy dynamics, e.g. quark-gluon interactions, are
integrated out of the theory. The resulting chiral effective
Lagrangian models the low-energy interactions between two or more
nucleons in terms of pion exchanges between nucleons and the
high-energy dynamics is incorporated as zero-ranged contact
interactions. This approach introduces several model parameters
referred to as low energy constants (LECs). They were denoted with
$\vec{\alpha}$ above, and play a central role during the model
calibration discussed below. The notion of high- and low-energy scales
in EFT requires the presence of at least two scales in the physical
system under study. An EFT formally exploits this scale separation to
expand observables in powers of the low-energy (soft) scale over the
high-energy (hard) scale, and in chiral EFT the resulting ratio is
often denoted
\begin{equation}
  Q = \frac{{\rm max}[m_{\pi},k]}{\Lambda_b}
\end{equation}
where, in the case of chiral EFT, the soft scales are $m_{\pi}$ and
$k$, the pion mass and a typical external momentum scale,
respectively. The hard scale is denoted $\Lambda_b$ and is set by the
e.g. the nucleon mass $M_N$. Depending on the system under study, one
can always try to exploit existing scale separations to construct
other kinds of EFTs in nuclear physics, e.g. pion-less
EFT~\cite{bedaque2002}, vibrational EFT~\cite{PAPENBROCK201136}, or
chiral perturbation theory (the prototypical EFT of
QCD)~\cite{GASSER1984142}. In the following, I will only discuss
results from \textit{ab initio} models based on chiral EFT, i.e. a
pion-full EFT, but many of the methods can be generally applied.

In chiral EFT, the nuclear interaction potential $V$ is analyzed as an
order-by-order expansion in terms of $Q^{\nu}$ and organized following
the principles of an underlying power counting (PC). Terms at a higher
chiral expansion-orders $\nu$ should be less important than terms at a
lower orders. Potentials expanded to higher orders are expected to
describe data better. Higher chiral orders contain more involved pion
exchanges and polynomial nucleon-contacts of increasing exponential
dimension, and therefore more undetermined model parameters
$\vec{\alpha}$ to handle during the calibration stage. To provide some
detail about the chiral potentials: the leading-order (LO) typically
consists of the familiar one-pion exchange interaction plus a
nucleonic contact-potential. The structure of the contact potential,
and the exact treatment of sub-leading orders vary depending on the
PC. Still, typical chiral potentials include at most contributions up
to a handful of chiral orders, e.g.  next-to-next-leading order (NNLO)
and next-to-next-to-next-to-leading order (N3LO), and the total number
of LECs, i.e. undetermined model parameters, range between $\sim
10-20$, sometimes a few more. Some of the unique advantages of chiral
EFT descriptions of the nuclear interaction are the natural emergence
of two-, three-, and many-nucleon
interactions~\cite{vankolck1994,epelbaum2002,PhysRevC.77.064004,PhysRevC.84.054001},
the consistent formulation of quantum currents, e.g. with respect to
electroweak
operators~\cite{PARK1993341,PARK1996515,KREBS2017317,PhysRevC.93.015501},
and a clear connection with the pion-nucleon Lagrangian which makes it
possible to link nuclei with low-energy pion-nucleon scattering
processes~\cite{HOFERICHTER20161}. For a detailed account of chiral
EFT potentials, see Refs.~\cite{vankolck1999, epelbaum2009,
  machleidt2011}.

To ensure steady progress towards a realistic \textit{ab initio} model
for atomic nuclei, we need to critically examine and evaluate the
quality and predictive power of different theoretical approaches and
model predictions. To this end it is crucial to equip all quantitative
theoretical results with uncertainties, and this is where another
advantageous aspect of EFT comes into play. It promises to deliver a
handle on the systematic uncertainty of a theoretical
prediction. Indeed, on a high level the EFT expansion for an
observable $\mathcal{O}$ can be written
\begin{equation}
  \mathcal{O} = \mathcal{O}_0 \sum_{\nu=0}^{\infty} c_{\nu} Q^{\nu},
  \label{eq:eft_expansion}
\end{equation}
where $\mathcal{O}_0$ is the first term in the above expansion, and
$c_{\nu}$ are dimensionless expansion coefficients. Here, and in the
following, the LO result ($\mathcal{O}_{0}$) was pulled out in front
of the sum to set the overall scale. One could equally well use the
experimental value for $\mathcal{O}$ or the highest-order calculation
to set the scale of the observable expansion. If we are dealing with
an EFT, one should expect the expansion coefficients to be of natural
size such that predictions at successive chiral orders are smaller by
a factor of $Q$. See also Refs.~\cite{lepage1997,grieshammer} for
discussions on how to assess the convergence of data. In an actual
calculation, the order-by-order description of $\mathcal{O}$ is
truncated at some finite order $k$, which induces a truncation error
$\delta_k$ in the prediction. The underlying EFT description then, in
principle, allows us to determine the formal structure of the
truncation error
\begin{equation}
  \delta_k = \mathcal{O}_0\sum_{\nu=k+1}^{\infty} c_{\nu}Q^{\nu}.
  \label{eq:eft_error}
\end{equation}
This type of handle on the theoretical uncertainty in a prediction is
not present in purely phenomenological descriptions of the nuclear
interaction such as the Argonne V18 potential~\cite{wiringa1995} or
the CD-Bonn potential~\cite{machleidt2001}. Despite all of the
promised advantages of chiral EFT, it should be pointed out that much
work remains to be done regarding the analysis and theoretical
underpinnings of chiral EFT, in particular the formulation of a PC
that, arguably, fulfills the field theoretic requirements for an EFT
of QCD, see e.g. Refs.~\cite{nogga2005, valderrama, long2012,
  Epelbaum2013, dyhdalo2016, sanchez2018, Epelbaum2018bb, yang2019}
for various views on this topic. Indeed, one cannot yet confidently
claim that the uncertainty estimates in \textit{ab initio} predictions
of nuclear observables based on proposed chiral EFT interactions are
linked to missing physics at the level of the effective
Lagrangian. The details of the PC, regularization approach, and chosen
maximum chiral order $k$ in Eq.~\ref{eq:eft_expansion}, are some of
many possible choices that give rise to the rich landscape of
different chiral interactions in nuclear theory. Although there is a
flurry of activity, and far from clear which is the best way to
proceed, there is tremendous overarching value to organize the model
analysis according to the fundamental ideas and expectations of EFT,
most importantly the promise of order-by-order improvement.

\section{Model Calibration}
\label{sec:model_calibration}
The goal of model calibration is to learn about the parameter of the
model using a pool of calibration data. This can mean many different
things depending on the situation, and in this section I will discuss
a few representative model calibration examples from \textit{ab
  initio} nuclear theory.

Assume that we have a model $M(\vec{\alpha};\vec{x})$ that consists of
a method for solving the Schr\"odinger equation and some theoretical
description of the nuclear interaction, e.g. a particular interaction
potential from chiral EFT, and we do not know the permissible values
for $\vec{\alpha}$. The vector $\vec{\alpha} = [\alpha_1,
  \alpha_2,\ldots,\alpha_N]$ denotes the $N$ physically relevant and
adjustable calibration parameters of model $M$, and the vector
$\vec{x}$ denotes the set of control inputs. The adjustable parameters
of interest will typically correspond to the LECs of the nuclear
interaction potential, and the vector $\vec{x}$ will contain
e.g. proton- and neutron-numbers, observable type, or some kinematical
setting. In principle the model might contain additional adjustable
parameters that for some reasons can be considered as constants. For
instance, we typically do not consider the pion mass as a calibration
parameter, although the variation of such fundamental properties can
also play an important role, see
e.g. Refs.~\cite{PhysRevC.76.054002,PhysRevLett.110.112502}.  The
choice of many-body method will depend on which class of observables
is targeted, either during prediction or calibration. For instance,
coupled-cluster theory will perform very well for nuclei in the
vicinity of closed shells and Faddeev integration will be able to
access the positive energy spectrum of the three-nucleon
Hamiltonian. Throughout, I will implicitly assume that the model is
realized only on a computer, i.e. $M$ is defined through some computer
code, and there is no stochastic element present in the output. This
means that each time the model is evaluated with the same input and
settings, we will basically get the same result.

To calibrate the parameters, suppose that we have a set of $n$
experimental observations compiled in a data vector $D =
[z_1,z_2,\ldots,z_n]$. They correspond to particular settings
$\vec{x}_1,\vec{x}_2,\ldots,\vec{x}_n$ of the control variables, to
produce model outputs for e.g. ground-state energies for light nuclei
or scattering cross sections at selected scattering momenta. We can
link the data points to the model outputs via the following relation
\begin{equation}
  z_i = M(\vec{\alpha},\vec{x}_i) + \delta(\vec{x}_i) + \varepsilon_i.
  \label{eq:data_model}
\end{equation}
This expression relates the reality of measurement with our model, and
includes a so-called model discrepancy term $\delta$, that depends on
the control variable $\vec{x}_i$. The measurement error is denoted
with $\varepsilon_i$. In cases where the measurement is accompanied
with \textit{zero} uncertainty, something that is highly unlikely of
course, the model discrepancy term represents the entire difference
between the model and reality. The theoretical discrepancy $\delta$ is
not physics \textit{per se}, but should rather be interpreted as a
random variable of statistical origin, informed via domain knowledge.

The model discrepancy term can be partitioned into at least three
terms
\begin{equation}
  \delta(\vec{x}_i) = \delta_{\rm interaction}(\vec{x}_i) + \delta_{\rm many-body}(\vec{x}_i) + \delta_{\rm numerical} (\vec{x}_i),
  \label{eq:discrepancy_decomposition}
\end{equation}
and they represent the neglected or missing physics in the theoretical
description of the nuclear interaction, neglected or missing many-body
correlations in the mathematical solution of the many-body
Schr\"odinger equation, and any numerical errors arising due to
algorithmic approximations in the implementation of the computer
model, respectively. We are currently most interested in understanding
$\delta_{\rm interaction}$ in situations where we, to a good
approximation, can neglect $\delta_{\rm many-body}$ and $\delta_{\rm
  numerical}$. Thus, in most of the literature, the dominant part of
the model discrepancy originates from the chiral EFT description of
the nuclear interaction. It should be pointed out that the discrepancy
term of the many-body method can be quite large for many types of
observables. However, \textit{ab initio} methods are often applied
wisely, and there exists plenty of domain knowledge regarding which
many-body methods that are best suited for different kinds of
observables. Yet, it is not easy to set bounds on this discrepancy
\textit{a priori}. Comparison between several complementary \textit{ab
  initio} models provides important
validation~\cite{PhysRevC.64.044001,hebeler2015c,Hergert_2016}. Finally,
the last term in Eq.~\ref{eq:discrepancy_decomposition} is currently
not the dominant part of the discrepancy, provided that the computer
code has been benchmarked.

Two related questions immediately arise: i) what is the impact of the
discrepancy term $\delta(\vec{x}_i)$ on the inference about the model
parameters $\vec{\alpha}$? and ii) what happens if we neglect
all sources of model discrepancy during model calibration?

Let us consider the second question, since it is easier and also sheds
light on the first one. Ignoring $\delta(\vec{x}_i)$ in
Eq.~\ref{eq:data_model} leaves us with the following expression
\begin{equation}
  z_i = M(\vec{\alpha},\vec{x}_i) + \varepsilon_i.
  \label{eq:data_model_no_err}
\end{equation}
This is the conventional starting point in nuclear model
calibration. If one also assumes that the measurement errors
$\varepsilon_i$ have finite variance, then the principle of maximum
entropy dictates that the likelihood of the data is normally
distributed. For independent errors, this leads to the canonical
expression for the likelihood
\begin{align}
  \begin{split}
  P(D|\vec{\alpha},M,\sigma) {}& = \prod_{i=1}^n \frac{1}{\sqrt{2\pi}\sigma_i} {\rm exp} \left\{-\frac{(z_i - M(\vec{\alpha},\vec{x}_i))^2}{2\sigma_i^2}\right \}\\
  {}& =\left[ \prod_{i=1}^n \frac{1}{\sqrt{2\pi}\sigma_i} \right]  {\rm exp} \left\{-\frac{1}{2}\sum_{j=1}^{n}\frac{(z_j - M(\vec{\alpha},x_j))^2}{\sigma_j^2}\right \} \\
  {}& = \left[ \prod_{i=1}^n \frac{1}{\sqrt{2\pi}\sigma_i} \right]  {\rm exp} \left\{-\frac{1}{2}\chi^2(\vec{\alpha})\right \}.
  \end{split}
  \label{eq:likelihood}
\end{align}
Here, the notation $P(X|Y)$ denotes the pdf of $X$ conditioned on
$Y$. The structure of the likelihood remains the same for correlated
measurement errors, although one must employ the full covariance
matrix instead of only the diagonal terms $\sigma_j^2$ to represent
the variance of the data. Model calibration in \textit{ab initio}
nuclear theory is typically formulated as a maximum likelihood
problem. This boils down to finding the optimal, or best-fitting,
parameters $\vec{\alpha}_{\star}$ that minimize the exponent in
Eq.~\ref{eq:likelihood}. We are thus facing a mathematical
optimization (minimization) problem
\begin{equation}
  \vec{\alpha}_{\star} = \underset{\vec{\alpha} \in \Omega}{\textnormal{arg min}}, \, \chi^2({\vec{\alpha}}),
\end{equation}
of finding the point that fulfills $\chi^2(\vec{\alpha}_{\star}) \leq
\chi^2(\vec{\alpha})$ for all $\vec{\alpha} \in \Omega$, where
$\Omega$ represents the parameter domain. In general, this is an
intractable problem unless we have detailed information about
$\vec{\alpha}$ or that the parameter domain is discrete and contains a
finite number of points. In reality, we are trying to find
\textit{local} minimizers to $\chi^2(\vec{\alpha})$, i.e. points
$\vec{\alpha}_{\star}$ for which $\chi^2(\vec{\alpha}_{\star}) \leq
\chi^2(\vec{\alpha})$ for all $\vec{\alpha} \in \Omega$ close to
$\vec{\alpha}_{\star}$.

For \textit{ab initio} models, optimization of the likelihood function
typically proceeds in several
steps~\cite{wiringa1995,machleidt2001,entem2003,carlsson2016,PhysRevC.91.024003,reinert2018}. First,
the parameters, i.e. the LECs in chiral EFT, are calibrated such that
the model optimally reproduces nucleon-nucleon scattering phase-shifts
from published partial-wave
analyses~\cite{stoks,PhysRevC.88.064002}. This typically yield model
parameters confined to some narrow range of values. Although each
scattering phase-shift only depends on a limited subset of the entire
vector of model parameters $\vec{\alpha}$, this stage still benefits
from using mathematical optimization algorithms, such as the
derivate-free algorithm called pounders~\cite{wild,ekstrom2013}. In a
next step, the results from the phase-shift optimization serves as the
starting point for a second round of parameter optimization where all
model parameters are varied to best reproduce thousands of
nucleon-nucleon scattering cross sections up to scattering energies in
the vicinity of the pion-production threshold.

Minimizing the $\chi^2$ in Eq.~\ref{eq:likelihood} for nucleon-nucleon
interaction potentials with respect to nucleon-nucleon scattering
data\footnote{A recent compilation of scattering data that is
  typically employed for this is provided in
  Ref.~\cite{PhysRevC.88.064002}.} has been the workhorse of model
calibration in nuclear theory for decades\footnote{The $\chi^2$
  function employed for nucleon-nucleon scattering data is slightly
  more involved to encompass partially correlated measurements, see
  e.g.~\cite{bergervoet}}. Since long, the figure of merit for a
nuclear interaction potential has been the $\chi^2$-per-datum
value. If this value is close to unity for some particular
parameterization $\vec{\alpha}_{\star}$, then the corresponding
potential is dubbed to be 'high-precision'. This is beginning to
change. Only for models $M$, where the model-discrepancy is in fact
negligible this approach can be justified. Otherwise, chasing a low
$\chi^2$ leads down the path of significant over-fitting, with
unreliable predictions as a consequence. For the calculation of
nucleon-nucleon scattering phase shifts and cross sections it is valid
to ignore $\delta_{\rm many-body}$ and $\delta_{\rm numerical}$ since
the corresponding equations are can be solved more or less numerically
exactly. However, since we clearly cannot claim to have a zero-valued
$\delta_{\rm interaction}$ term, the $\chi^2$-per-datum with respect
to nucleon-nucleon scattering data is not the optimal measure to guide
future efforts in nuclear theory. Before and during the development of
\textit{ab initio} many-body methods and EFT principles, when it was
very unclear how to understand the concept of model discrepancy in
nuclear theory, it was certainly more warranted to benchmark nuclear
potentials based solely on a straightforward $\chi^2$ value.

State-of-the-art interaction potentials also contain three-nucleon
force terms. Although some of the parameters in chiral EFT are shared
between two- and three-nucleon terms, there exists a subset of
parameters inherent only to the three-nucleon interaction. Such
parameters must be determined using observables from $A>2$
systems. Arguably, all parameters of a chiral potential should be
optimized simultaneously to a joint dataset $D$. The parameters must
therefore be informed about e.g. binding energies and charge radii of
$^{3,4}$He and $^{3}$H. Unfortunately, there exists a universal
correlation between the binding energies of $^{3}$H and $^{4}$He, the
so-called Tjon line~\cite{tjon}, which reduces the information content
of this data set. Ground state-energies and radii are also highly
correlated. Fortunately, it was demonstrated in Ref.~\cite{gazit2009}
that the beta decay of $^{3}$H can add valuable information about the
parameters in the three-nucleon interaction. Most methods for solving
the $A=2,3,4$ Schr\"odinger equation for bound states do so with
nearly zero many-body discrepancy. Recently, selected three-nucleon
scattering observables have been added to the pool of calibration
data~\cite{PhysRevC.99.024313}, however not routinely since it is
still computationally quite costly to evaluate the \textit{ab initio}
models for such observables. There are indications that it is
necessary to include also data from nuclei heavier than $^{4}$He to
learn about the parameters in \textit{ab initio} models. This is
discussed in Sec.~\ref{sec:data}.

Ignoring the $\delta_{\rm interaction}$ discrepancy terms during model
calibration can have serious consequences. Most importantly, this
reduces the LECs to tuning parameters without \textit{any} physical
meaning. Indeed, in the strive to replicate the data at any cost, the
numerical values can be driven far away from the \textit{true} values
of the model. At some point, continued tuning of the parameters
induces over-fitting and the model will pick up on the noise in the
data. Naturally , this leads to poor predictive power. With increasing
amounts of data, the optimization process will converge with
increasing certainty to false values for $\vec{\alpha}$. A pedagogical
introduction to the statistics of model discrepancies and a physics
example is provided in Ref.~\cite{brynjarsdottir}.

A total model discrepancy, according to
Eq.~\ref{eq:discrepancy_decomposition}, was included in \textit{ab
  initio} model calibration for the first time in
Ref.~\cite{carlsson2016}. The parameters in a set of chiral
interactions at LO, NLO, and NNLO were optimized using
nucleon-nucleon, and pion-nucleon scattering data. The terms in the
three-nucleon interaction were simultaneously informed using
bound-state observables from $A=2,3$ nuclei. The details of the
analysis and results can be found in the original paper. The
discrepancy terms were interpreted as uncorrelated errors and added in
quadrature with the data uncertainties, leading to a slight
modification of the corresponding $\chi^2$ function
\begin{equation}
  \chi^2 = \sum_{j=1}^{n}\frac{(z_j -
    M(\vec{\alpha},x_j))^2}{\sigma_{{\rm data},}j^2 + \sigma^2_{{\rm
        interaction},j} + \sigma^2_{{\rm many-body},j} +
    \sigma^2_{{\rm numerical},j}}.
  \label{eq:chi2_modified}
\end{equation}
The interaction discrepancy was constructed from the EFT assumption
that the external momenta flowing through the interaction diagrams
scale as some power corresponding to the truncation of the chiral
expansion, in accordance with Eq.~\ref{eq:eft_error}. The intrinsic
scale of this error was solved for self-consistently by requiring that
the $\chi^2$-per-datum should approach unity providing that the model
error is correctly estimated. This implicitly assumes a correct
estimate of the number of statistical degrees of freedom. Something
that cannot be easily estimated for non-linear $\chi^2$
functions~\cite{andrae2010dos}.

To summarize, although the inclusion of model discrepancies is
preferred, it is not without problems. To blindly include a term
$\delta(\vec{x}_i)$ to capture model discrepancies in the process of
model calibration can lead to statistical confounding between
$\vec{\alpha}$ and $\delta(\cdot)$~\cite{brynjarsdottir}. This means
that the model parameters and the discrepancy term are not
identifiable and we only recover a some joint pdf for the two
components. Indeed, for any $\vec{\alpha}$ there is a $\delta(\cdot)$
given by the difference between model and reality. To make progress
requires us to specify some \textit{a priori} ranges for
$\vec{\alpha}$ and/or $\delta(\cdot)$. Or in the language of Bayesian
inference, we need to specify the prior pdf for the model parameters
and the theory uncertainties. This is partly related to approaches
where one augments the $\chi^2$ function with a penalty term to
constrain the values of the model parameters, see
e.g. Ref.~\cite{Furnstahl_2015}. For EFT descriptions of the nuclear
interaction one can argue that the LECs should maintain values of
order unity, if expressed in units of the breakdown scale, and the
discrepancy could follow the pattern of Eq.~\ref{eq:eft_error}. To
adequately represent the discrepancy term in nuclear models is ongoing
research , and it appears advantageous to reformulate model
calibration as a Bayesian inference problem, see Sec.~\ref{sec:bayes}.
\subsection{Hessian error analysis}
\begin{figure}[t]
\begin{center}
\includegraphics[width=13cm]{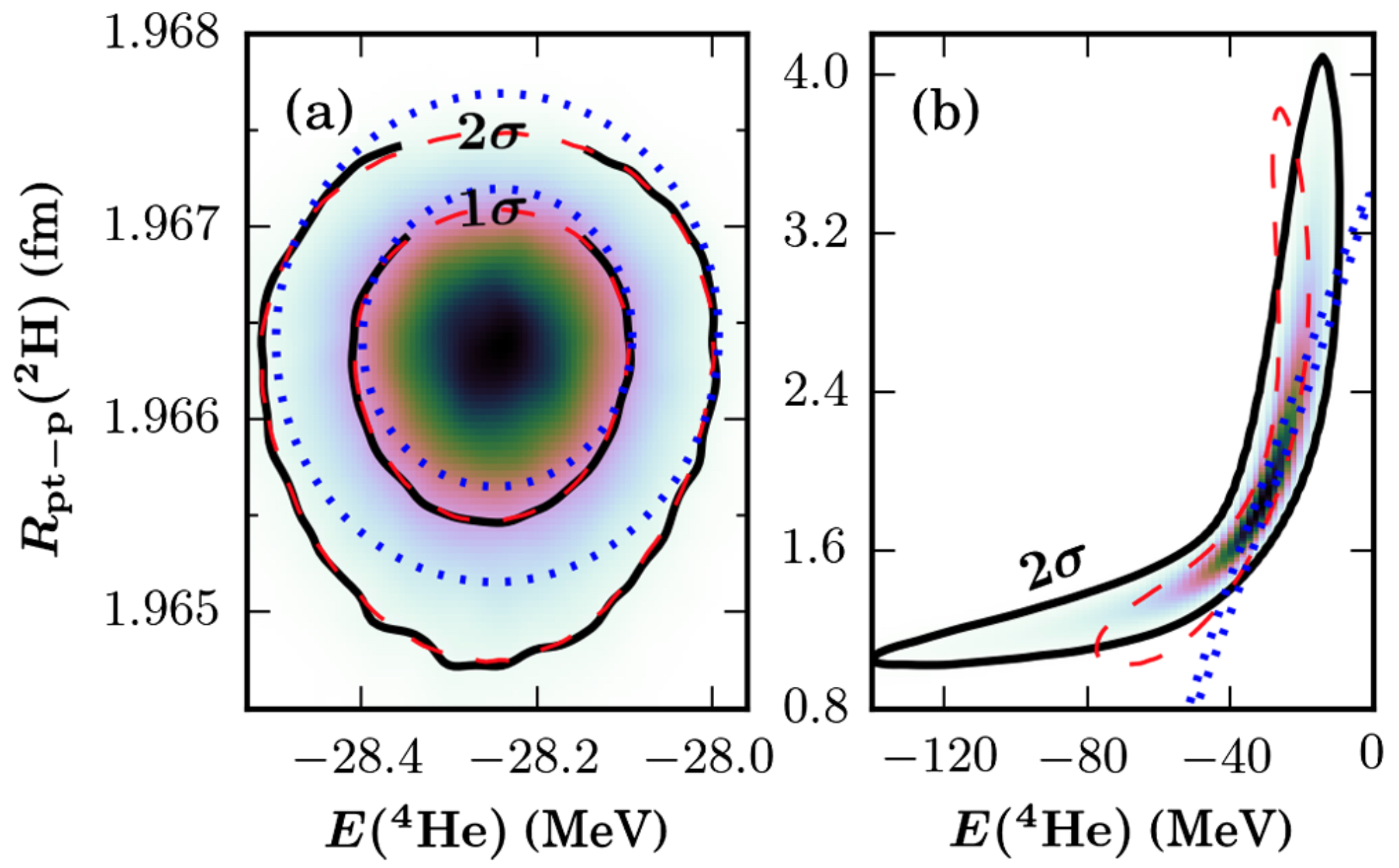}
\end{center}
\caption{ Joint distribution for the ground-state energy of $^4$He
  (x-axis) and the point-proton radius of $^{2}$H (y-axis) for (a) the
  chiral potential NNLO$_{\rm sim}$ and (b) the chiral potential
  NNLO$_{\rm sep}$, see Ref.~~\cite{carlsson2016}. Contour lines for
  the distributions are shown as black solid lines, while blue dotted
  (red dashed) contours are obtained assuming a linear (quadratic)
  dependence on the LECs for the observables.}\label{fig:4He_2H}
\end{figure}

At the optimum parameter point $\vec{\alpha}_{\star}$, a
Taylor expansion of the $\chi^2$ function to second order gives
\begin{align}
  \begin{split}
    \chi^2(\vec{\alpha}_{\star} + \Delta \vec{\alpha}) {}& \approx \chi^{2}(\vec{\alpha}_{\star}) + \frac{1}{2}(\Delta \vec{\alpha})^{T} H (\Delta \vec{\alpha}), \\
{}& {\rm where} \,\,\,    \left. H_{ij} = \frac{\partial^2 \chi^2 (\vec{\alpha})}{\partial \alpha_i \partial \alpha_j} \right|_{\vec{\alpha} = \vec{\alpha}_{\star}},
  \end{split}
\end{align}
where $H$ denotes a Hessian matrix, the inverse of which is
proportional to the covariance matrix for the model
parameters~\cite{Dobaczewski_2014}. Contracting the parameter-Jacobian
of any model prediction with this covariance matrix yields the
standard error propagation result of the parameter uncertainties. For
the conventional $\chi^2$ function, the parameter covariances reflect
the impact of the experimental uncertainties on the precision of the
optimum and predicted observables. Sometimes, this is referred to as
statistical uncertainties, which is a bit confusing since \textit{all}
uncertainties are statistical in nature. See Fig.~\ref{fig:4He_2H} for
an example result of applying a parameter covariance matrix to obtain
the joint pdf for the $^{4}$He ground-state energy and the $^{2}$H
point-proton radius, two important few-nucleon observables. This
particular result is taken from Ref.~\cite{carlsson2016}, where in
fact a model discrepancy term $\delta(\cdot)$ was incorporated during
the optimization, thus in this particular case the covariances reflect
more than just the measurement noise. See
e.g. Refs.~\cite{ekstrom2015b, PhysRevC.92.064003, PhysRevC.89.064006,
  acharya2016, Acharya2018, HERNANDEZ2018377} for details about
statistical error analysis and illuminating examples of forward error
propagation in \textit{ab initio} nuclear theory.

To extract the covariance matrix requires computation of the
second-order derivatives of the $\chi^2$ function with respect to the
model parameters. The general process of numerically differentiating
an \textit{ab initio} model with respect to $\vec{\alpha}$ is
significantly simplified, and numerically much more precise, with the
use of automatic differentiation (AD)~\cite{carlsson2016}. This
corresponds to applying the chain rule of differentiation on a
function represented as a computer code. It relies on the principle
that any computer code, no matter how complicated always executes a
set of elementary arithmetic operations on a finite set of elementary
functions (exponentiation, logarithmization, etc). To implement AD
requires modification of the original computer code, e.g. operator
overloading via third-party libraries. Once implemented, AD also
enables application of more advanced derivative-based optimization
algorithms and Markov chain Monte Carlo methods with the computer
model $M$. An alternative, and derivative-free approach, to computing
the Hessian matrix for forward error propagation is to employ Lagrange
multipliers~\cite{Carlsson2017}. This method is more robust, but also
more computationally demanding to carry out. From a practical and
computational perspective, if one considers to use Lagrange
multipliers, one should also look into performing a Bayesian analysis,
see Sec.~\ref{sec:bayes}.

\subsection{Selecting calibration data}
\label{sec:data}
It is preferable to use data corresponding to observables that are
computationally cheap to evaluate, and if possible with model settings
corresponding to low $\delta(\vec{x})_{\rm many-body}$
discrepancies. One should also strive to include data with highly
complementary information content that constrain a maximum amount of
linearly independent combinations of model parameters.

\begin{figure}[t]
\begin{center}
\includegraphics[width=13cm]{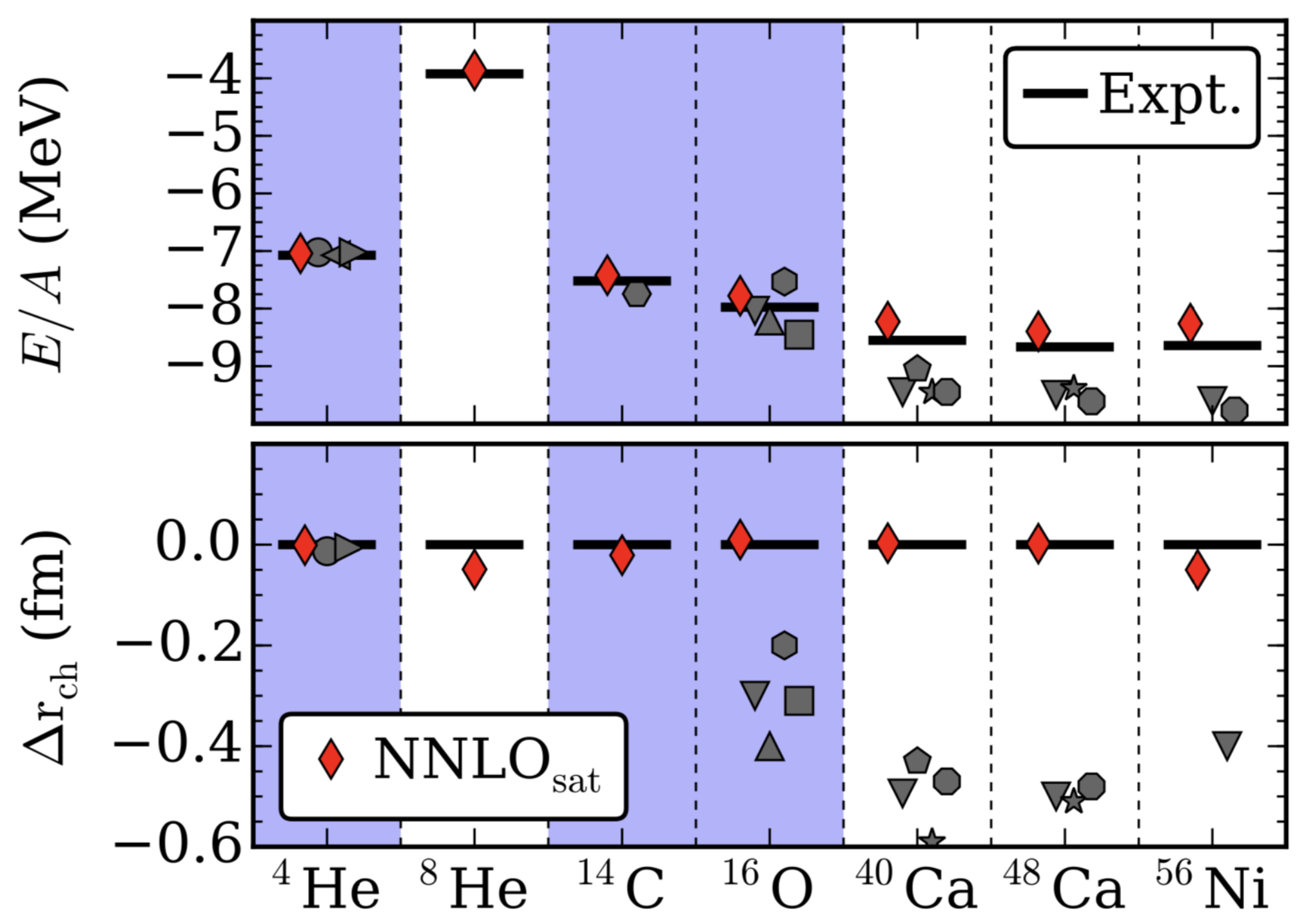}
\end{center}
\caption{Ground-state energies per nucleon (top) and differences
  between theoretical and experimental charge radii (bottom) for
  selected light and medium-mass nuclei and results from ab initio
  computations. The red diamonds mark results based on the chiral
  interaction NNLO$_{\rm sat}$. The blue columns indicate which nuclei
  where included in the optimization of the LECs in NNLO$_{\rm sat}$,
  while the white columns are predictions. Grey symbols indicate other
  chiral interactions. }
\label{fig:NNLOsat}
\end{figure}

The conventional approach to calibrate \textit{ab initio} models is to
use only data from $A \lesssim 4$ nuclei, as was discussed above. It
was observed in Ref.~\cite{ekstrom2015} that the additional inclusion
of ground-state energies and charge radii of selected carbon and
oxygen isotopes dramatically increases the predictive power of models
for bulk properties of nuclei up to the medium-mass nickel region, see
Fig,~\ref{fig:NNLOsat}. This calibration strategy led to the
construction of the so-called NNLO$_{\rm sat}$ interaction. From a
quantitative perspective, the advent of models capable of accurate
predictions is of course an important step forward and has proven very
useful~\cite{hagen2015,garciaruiz2016,PhysRevC.98.044625,PhysRevLett.123.092501}.

The major drawback of any model based on the NNLO$_{\rm sat}$
interaction is the lack of quantified theoretical uncertainties. This
is quite common also for \textit{ab initio} models based on other
interaction potentials. At the moment, the best we can do is to
estimate the truncation error using Eq.~\ref{eq:eft_error}. This
requires additional and sub-leading chiral-order potentials using the
same optimization protocol, e.g. LO$_{\rm sat}$ and NLO$_{\rm sat}$,
which do not exist. The calibration of such models require an even
more careful inclusion of model discrepancies. This is discussed more
in Sec.~\ref{sec:bayes}. One can certainly argue that it becomes even
more important to quantify the theory errors for models that we
strongly believe will make accurate predictions, like the ones based
on the NNLO$_{\rm sat}$ interaction. Otherwise we are limited in our
ability to assess discrepancies with respect to experiment. This
argument applies equally well to models based on e.g. the 1.8/2.0
interaction from Ref.~\cite{nogga2004,hebeler2011} which typically
yield good predictions for binding energies and low-energy spectra. In
Ref.~\cite{hagen2015}, the prediction from \textit{ab initio} models
based on different interactions, NNLO$_{\rm sat}$ and the 1.8/2.0
interactions amongst other, were compared to estimate the overall
theoretical uncertainty.

It is difficult to judge the degree of over-fitting to finite nuclei
in NNLO$_{\rm sat}$. It was noted during calibration that this
interaction fails to reproduce experimental nucleon-nucleon scattering
cross sections for scattering momenta larger than $\sim m_{\pi}$.
Enforcing a good reproduction of all scattering data up to e.g. the
pion-production threshold most likely corresponds to over-fitting in
the $A=2$ sector. It is the role of the model discrepancy term, with
appropriate priors, to balance this.

One clearly gains predictive power by including additional medium-mass
data during the calibration stage. This was also observed in a lattice
EFT analysis of the nuclear binding
mechanism~\cite{PhysRevLett.117.132501}. The related topic of possibly
emergent nuclear phenomena like saturation, binding, and deformation
of atomic nuclei is discussed further in
Ref.~\cite{hagen2016}. Although the inclusion of a model discrepancy
term while calibrating to heavier-mass data will be important, it does
not solve the underlying problem of having a systematically uncertain
model. It was noted in
Refs.~\cite{ekstrom2018,piarulli2018,logoteta2016} that the explicit
inclusion of the $\Delta$ isobar in the chiral description of the
nuclear interaction dramatically improves the description of nuclei
while also reproducing nucleon-nucleon scattering data. A possibly
fruitful way forward is to employ improved models, i.e. with explicit
inclusion of the $\Delta$ isobar, that are calibrated using also data
from selected heavy-mass nuclei, while systematically accounting for
model discrepancies. Furthermore, it will be interesting to se how
much additional information is contained in three-nucleon scattering
data~\cite{Kalantar_Nayestanaki_2011}.

\section{Bayesian inference}
\label{sec:bayes}
The previous section introduced the concept of model calibration and
the fundamental expression in Eq.~\ref{eq:data_model} that relates a
model with measured data. In this section I will outline the Bayesian
strategy for learning about the model parameters and some existing
estimates of the discrepancy term. The overarching goal is still to
calibrate an \textit{ab initio} model $M(\vec{\alpha},\vec{x})$, and
reliably predict properties of atomic nuclei. However, instead of
finding a single point $\vec{\alpha}_{\star}$ in parameter space that
maximizes the likelihood for the data, we can use Bayes' theorem to
relate the data likelihood to a pdf for the model parameters
themselves
\begin{equation}
  P(\vec{\alpha}|D,M,I) = \frac{P(D|\vec{\alpha},M,I)P(\vec{\alpha}|M,I)}{P(D|M,I)},
  \label{eq:bayes}
\end{equation}
where $P(\vec{\alpha}|M,I)$ denotes the prior pdf for the parameters,
$P(D|\vec{\alpha},M,I)$ denotes the likelihood of the data, the
denominator $P(D|M,I)$ denotes the marginal likelihood of the data,
and $P(\vec{\alpha}|D,M,I)$ denotes the sought-after posterior pdf of
the model parameters. The additional $I$ represents any other
information at hand.

The Bayesian reformulation of the inference problem can at first sight
appear as a subtle point, and it is easy to overlook the fundamental
difference between computing the pdf for the parameters and maximizing
the likelihood, i.e. frequentist inference. From a practical
perspective, it is clearly advantageous to obtain a pdf for the model
parameters $P(\vec{\alpha}|D,M)$. This quantity is also intuitively
straightforward to interpret compared to frequentist interval
estimates that might contain the true value of the unknown model
parameters, e.g. confidence intervals. The prior pdf
$P(\vec{\alpha}|M,I)$ for the parameters $\vec{\alpha}$ given a model
$M$ offers up front possibility to incorporate any prior knowledge (or
belief) about the parameters, before we look at the data. In the case
of \textit{ab initio} modelling, an underlying EFT-description of the
nuclear interaction embodies substantial prior knowledge, such as the
typical magnitude of the model parameters as well as a handle on the
systematic uncertainty. The Bayesian requirement of prior
specification also ensures full transparency regarding the
assumptions that goes into the analysis.

The existence of priors in Bayesian inference is sometimes criticized
and one can argue that the scientific method should let the data speak
for itself, without the explicit insertion of subjective prior
belief. Inference about model parameters in terms of hypothesis tests
or confidence intervals, derived from the frequency of the data, is
referred to as frequentist inference. Note however that the likelihood
rests on initial subjective choice(s) regarding the data model. In
this review, I will maintain a practical perspective, and just
recognize the usefulness of the Bayesian approach to encode prior
information about the model parameters and the model discrepancy
terms. Which is also required in order to handle possible confounding
between the discrepancy and the model
parameters~\cite{brynjarsdottir}. Either way, it is difficult to avoid
subjective choices in statistical inference involving uncertainties
and limited data. In fact, one can even argue that only
\textit{subjective} probabilities exist~\cite{definetti}.

Bayesian model calibration, sometimes called Bayesian parameter
estimation, is currently emerging in \textit{ab initio}
modelling~\cite{schindler2009,Wesolowski2015,Wesolowski2019}. To get
get some intuition about this topic, let us look at Bayesian parameter
estimation in its most simple version. This amount to assuming a
(bounded) uniform prior pdf for the model parameters $\vec{\alpha}$,
i.e.
\begin{equation}
  P(\vec{\alpha}|M,I) \sim \mathcal{U}(\vec{a},\vec{b})
\end{equation}
and adopting a data likelihood as in Eq.~\ref{eq:likelihood}. In
practice, what remains is to explicitly evaluate
$P(\vec{\alpha}|D,M,I)$ in Eq.~\ref{eq:bayes} by computing the product
of the two terms in the numerator.  The denominator can be neglected
since it does not explicitly depend on $\vec{\alpha}$. This marginal
likelihood does however matter for absolute normalization of the
posterior pdf. The evaluation of the posterior can be done via brute
force evaluation in some simple cases, but for computationally
expensive models and/or high-dimensional parameter space typically
more clever strategies are required, such as Markov chain Monte
Carlo. With uniform priors, the point for the maximum posterior
coincides exactly with the point obtained using maximum likelihood
methods, which for normal likelihood distributions is nothing but
least-squares. 

The advantages of Bayesian parameter estimations becomes apparent once
we include non-uniform prior knowledge, and in most cases we know a
bit more about the parameters than what a simple uniform pdf
reflects. The general strategies for application of Bayesian methods
to calibrate EFTs are pedagogically outlined in
Ref.~\cite{Wesolowski2015}. To exemplify the use of priors and some of
the related techniques, let us assume a Gaussian prior with zero mean
for the model parameters $\vec{\alpha} =
[\alpha_1,\alpha_2,\ldots,\alpha_N]$, i.e.
\begin{equation}
  P(\vec{\alpha}|\bar{a},M,I) = \left( \frac{1}{\sqrt{2\pi} \vec{a}}\right)^{N} {\rm exp} \left( -\frac{\vec{\alpha}^2}{2\bar{a}^2}\right),
\end{equation}
where the parameter $\bar{a}^2$ denotes the prior variance. This is
not an unreasonable prior for the model parameters in chiral EFT. The
impact of this parameter prior is to penalize model parameters that are
too large, which would typically signal over-fitting. For situations
where there exist a large amount of precise data, the prior
specification for the parameters matter less. Nevertheless, the
question remains, what value should we pick for $\bar{a}$? This
can be dealt with straightforwardly by marginalizing over
$\bar{a}$, i.e. we express the prior for the parameters as
\begin{equation}
  P(\vec{\alpha}|M,I) = \int {\rm d}\bar{a} \, P(\vec{\alpha}|\bar{a},M,I)P(\bar{a}|M,I),
\end{equation}
which only forces us to specify a prior for the variance for our
belief about the model parameters, here we could choose a rather broad
range if we like. With appropriate analytical form for the prior on
$\bar{a}$, it is even possible to carry out this marginalization step
analytically. See Ref~\cite{Wesolowski2019} for illuminating examples
about the impact of different priors in model calibration with
scattering-phase shifts.

\subsection{Prediction and calibration including model discrepancies}
Observables computed with potentials from chiral EFT should exhibit a
pattern where contributions from successive orders
$\nu=0,1,2,3,\ldots$ are smaller by factors $Q^{\nu}$. This is
reflected in Eq.~\ref{eq:eft_expansion}. Therefore the expansion
coefficients $\{ c_{\nu}\}$ should remain of natural size, a clear
example of a situation where we have prior knowledge\footnote{The
  wording; prior knowledge vs. prior expectation, or even prior
  belief, signals the level of subjective certainty or source for the
  prior.}. Given a series of model calculations of the observable
$\mathcal{O}$, up to the chiral order $\nu=k$,
i.e. $\mathcal{O}_{0},\mathcal{O}_{1},\ldots,\mathcal{O}_{k}$, and an
estimate of the factor $Q$, it is straightforward to extract the
coefficients $[c_0,c_1,\ldots,c_{k}]$. It was shown in
Refs.~\cite{Cacciari2011,furnstahl2015} how to extract a pdf for the
EFT truncation error $\delta_k$ in Eq.~\ref{eq:eft_error} using this
information. First, we factor out the overall scale, and define
\begin{equation}
  \tilde{\delta}_{k} = \delta_k/\mathcal{O}_0 
\end{equation}
as the overall dimensionless truncation error. We now seek an
expression for $P(\tilde{\delta}_k|c_{0},c_1,\ldots,c_k)$ given
the known values for the first $k+1$ coefficients. It turns out that
for independent, bounded, and uniform prior pdfs for the expansion
coefficients, the integrals can be solved analytically if one also
approximates $\tilde{\delta}_{k}$ with the leading term. Thus, we
assume
\begin{equation}
  \tilde{\delta}_{k} \approx \tilde{\delta}_k^{(1)} = c_{k+1}Q^{k+1},
\end{equation}
The posterior pdf $P(\tilde{\delta}_k^{(1)}|c_{0},c_1,\ldots,c_k)$ is
given in Ref.~\cite{furnstahl2015} [Eq. 22], and explicitly derived in
the appendix of Ref.~\cite{Cacciari2011}. This posterior pdf is the
complete inference about $\tilde{\delta}_{k}^{(1)}$. If the pdf is
multi-modal or otherwise non-trivial one should use it in its entirety
in forward analyses. However, we can sometimes use a so-called degree
of belief (DOB) value to quantify the width of a pdf. This is the
probability $p\%$, expressed in percent, that the value of an
uncertain variable $\eta$, distributed according to the pdf $P(\eta)$, falls
within an interval $[a,b]$. This interval is then referred to as a
credible interval with $p\%$ DOB, where
\begin{equation}
  p\% = \int_{a}^{b} P(\eta) \,\, {\rm d}\eta.
\end{equation}
The posterior pdf for $\tilde{\delta}_{k}^{(1)}$ is not Gaussian,
however it is symmetric and have zero mean. Therefore, we can define a
smallest interval $[-d_k^{(p)},+d_k^{(p)}]$ that captures $p\%$ of the probability mass
\begin{equation}
  p\% = \int_{-d_k^{(p)}}^{+d_k^{(p)}}  P(\tilde{\delta}_{k}^{1}| c_{0},c_1,\ldots,c_k) \,\, {\rm d}\tilde{\delta}_{k}^{(1)},
  \end{equation}
and solve for $d_{k}^{(p)}$. This will define the width of the
credible interval within which the next term in the EFT expansion will
fall with $p\%$ DOB, i.e. an estimate of the truncation
error. The expression is derived
in~Refs.~\cite{Cacciari2011,furnstahl2015}, and given by
\begin{equation}
  d_{k}^{(p)} = {\rm max}(|c_0|,|c_1|,\ldots,|c_k|)Q^{k+1}
  \frac{n_c+1}{n_c} p\% \,\,\, ,{\rm if} \,\, p\% \leq n_c/(n_c+1),
  \label{eq:eft_dob}
\end{equation}
 where $n_c$ denotes the number of available coefficients. Thus, with
 $n_c/(n_c+1) \times 100\%$ DOB, the EFT truncation error for the
 observable $\mathcal{O}$, in dimensionful units, is straightforwardly
 estimated by $\mathcal{O}_0 \times {\rm
   max}(|c_0|,|c_1|,\ldots,|c_k|)Q^{k+1}$. This estimate also
 corresponds to the prescription employed in
 Ref.~\cite{Epelbaum2015b}.  This \textit{a posteriori} truncation
 error estimate essentially boils down to guessing the largest number
 that one can expect based on a series of numbers drawn from the same
 underlying distribution. For example, given only one ($n_c=1$)
 expansion parameter $c_{0}$, we have a $50\%$ DOB that we have
 encountered the largest coefficient in the series. This procedure has
 been applied to estimate the truncation error in several \textit{ab
   initio} model calculations, see the long list of papers that are
 citing Refs.~\cite{furnstahl2015,Epelbaum2015b}.
\begin{figure}[th]
\begin{center}
  \includegraphics[width=18cm]{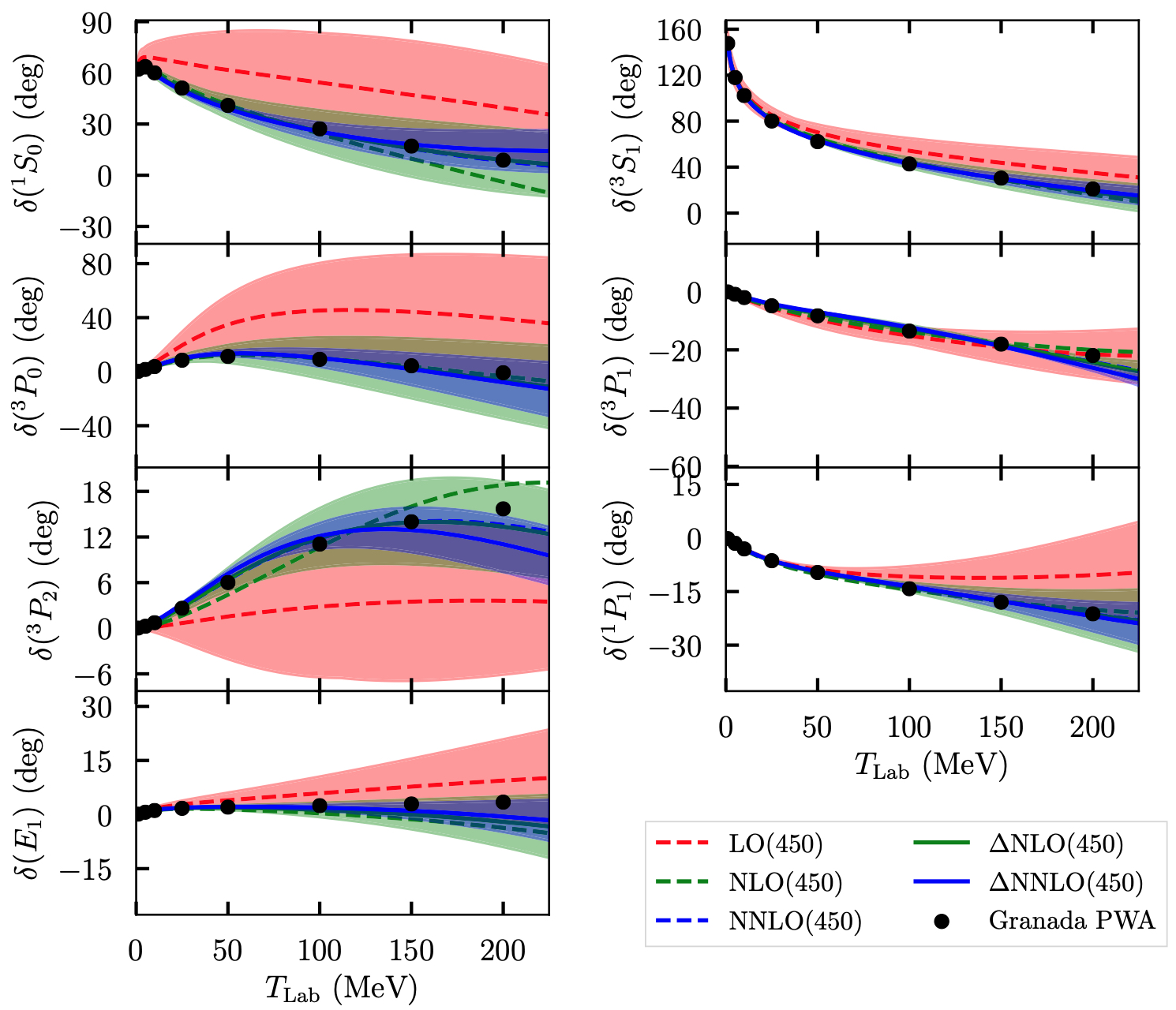}
\end{center}
\caption{Neutron-proton scattering phase shifts computed with models
  based on $\Delta$-full and $\Delta$-less chiral interaction
  potentials. The bands indicate the limits of the expected DOB
  intervals at each chiral order $\nu$. The black dots represent the
  values from the Granada partial wave
  analysis~\cite{PhysRevC.88.064002}.}\label{fig:phase_shifts}
\end{figure}

\begin{figure}[t]
\begin{center}
  \includegraphics[width=12cm]{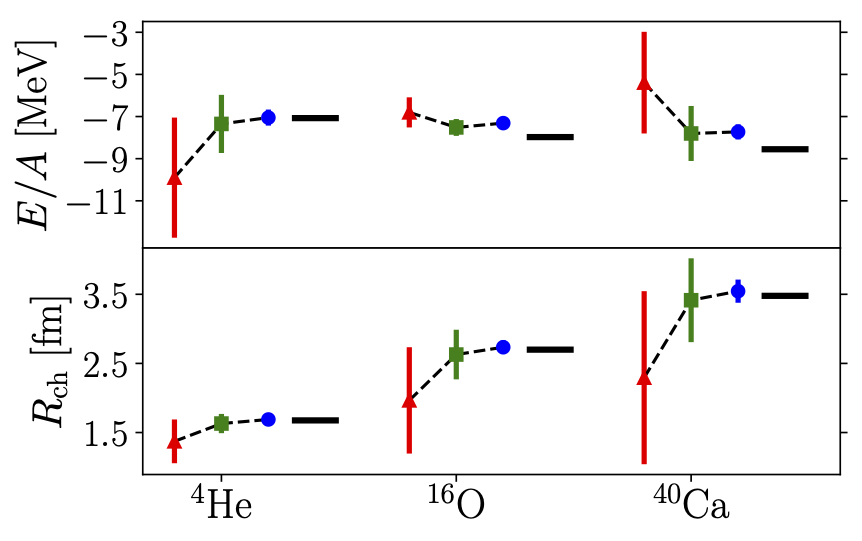}
\end{center}
\caption{Ground-state energy (negative of binding energy) per nucleon
  and charge radii for selected nuclei computed with coupled cluster
  theory and the $\Delta$-full potential $\Delta$NNLO(450). For each
  nucleus, from left to right as follows: LO (red triangle), NLO
  (green square), and NNLO (blue circle). The black horizontal bars
  are data. Vertical bars estimate uncertainties from the
  order-by-order EFT truncation errors.}\label{fig:finite}
\end{figure}

\begin{figure}[t]
\begin{center}
  \includegraphics[width=12cm]{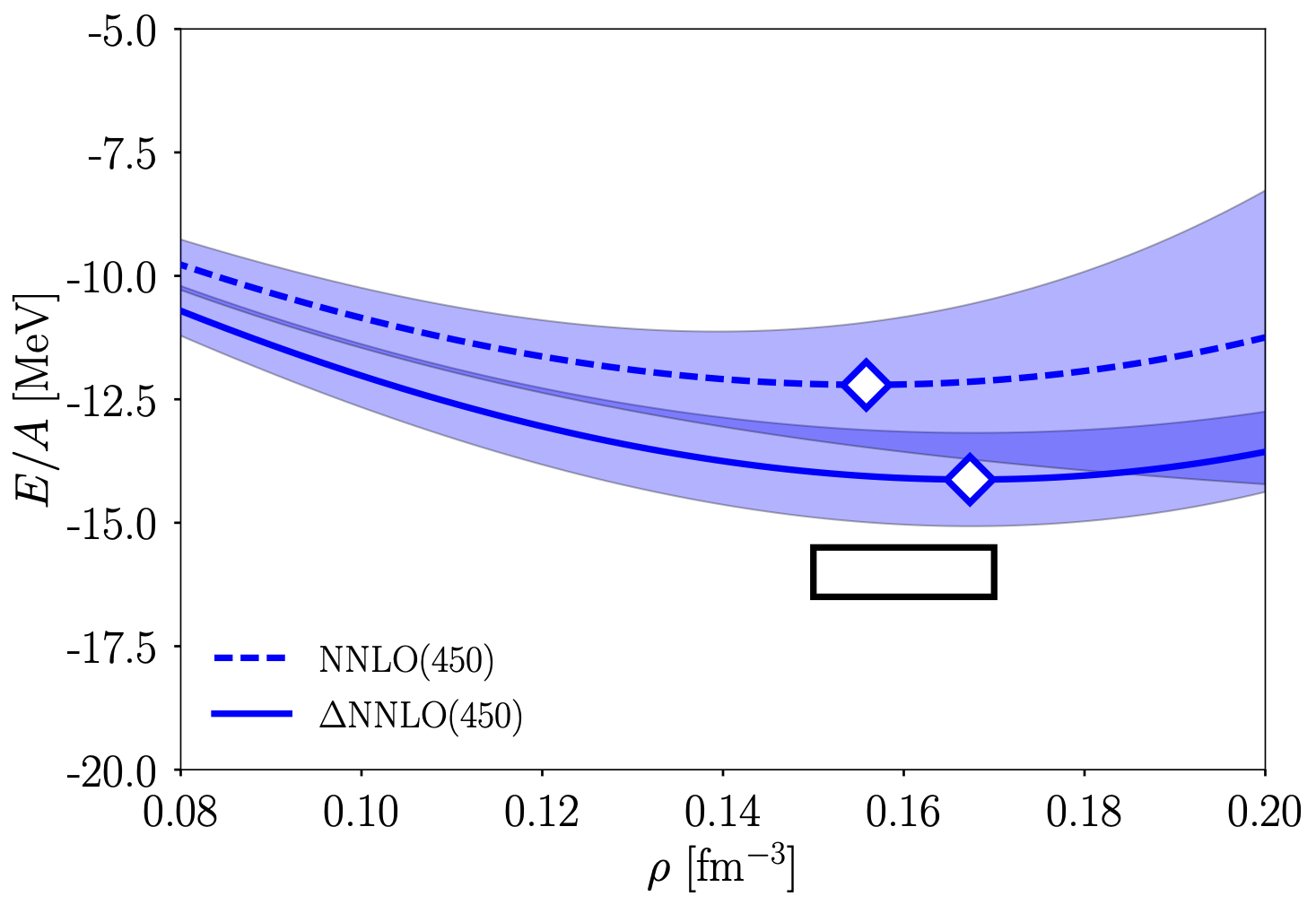}
\end{center}
\caption{Coupled-cluster based model prediction of the energy per
  nucleon (in MeV) in symmetric nuclear matter using an NNLO potential
  with (solid line) and without (dashed line) the $\Delta$
  isobar. Both interactions employ a momentum regulator-cutoff
  $\Lambda$ = 450 MeV. The shaded areas indicate the estimated
  EFT-truncation errors. The diamonds mark the saturation point and
  the black rectangle indicates the region $E/A$ = -16 $\pm$ 0.5 MeV
  and $\rho$ = 0.16 $\pm$ 0.01 fm$^{-3}$.}\label{fig:matter}
\end{figure}

The procedure for estimating the EFT truncation error, i.e. part of
the model discrepancy, requires an estimate of the high-energy scale
$\Lambda_b$ of the underlying EFT. For the models discussed here, the
results are based on chiral EFT, for which the naive estimate of
$\Lambda_b$ is roughly $M_N\sim$1 GeV. This was analyzed more
carefully for semi-local chiral
potentials~\cite{epelbaum2015,Epelbaum2015b} in
Ref.~\cite{PhysRevC.96.024003}. The posterior pdf for $\Lambda_b$
indicated that a more probable value is $\Lambda_b \approx 500$
MeV. This value was also used for the breakdown scale in the
truncation error analysis of nucleon-nucleon scattering phase shifts
from the $\Delta$-full models at LO,NLO, and NNLO chiral orders in
Ref.~\cite{ekstrom2018}. The results are presented in
Fig.~\ref{fig:phase_shifts}. This result also strengthens the
observation made earlier, that the inclusion of the $\Delta$ degree of
freedom tend to improve model descriptions of nuclear systems. This is
more clearly seen when employing the same potentials to make model
predictions for the ground-state-energies and charge radii of selected
finite nuclei, see Fig.~\ref{fig:finite}, and the energy per nucleon
in symmetric nuclear matter, see Fig.~\ref{fig:matter}.
The model predictions for the nuclear matter indicate that the
$\Delta$-full models on average agree better with experimental
energies and radii. The uncertainty bands for the predictions were
extracted under the additional assumption that the relevant
soft-scales for finite and infinite nuclear systems are given by the
pion mass and the Fermi momentum, respectively. Although these are
rough estimates of the soft scales, it is important to note that the
the truncation error in Eq.~\ref{eq:eft_dob} only holds up to factors
of order unity. A comparison of theoretical error estimates based on
different statistical methods provide additional validation. The
Bayesian method for estimating the truncation error and the model
errors estimated using the modified $\chi^2$-function in
Eq.~\ref{eq:chi2_modified} are quite different in
nature. Nevertheless, a comparison of the theoretical errors in
nucleon-nucleon cross sections at high scattering-energies agree very
well for these methods~\cite{carlsson2016,furnstahl2015}. The link
between the two approaches for estimating the model uncertainties is
discussed further in Ref.~\cite{Wesolowski2019}. A complete Bayesian
parameter estimation including model discrepancy will hopefully reveal
more details about the structure of the chiral EFT error.

At the moment, most model discrepancies in \textit{ab initio}
modelling based on chiral EFT are extracted \textit{a posteriori}
using predictions based on calibrated models. This is possible based
on the expectation that the predictions might follow an EFT
pattern. This of course remains to be validated on theoretical
grounds. However, under the assumption that the interaction potential
actually gives rise to an EFT pattern for the observable, we can build
on Eq.~\ref{eq:eft_error} to include a discrepancy term in the
likelihood for calibrating \textit{ab initio} models. See
Ref.~\cite{PhysRevC.100.044001} for a discussion about correlated
truncation errors in nucleon-nucleon scattering observables following
this line of thought, where it is also observed that the expansion
parameters behave largely as expected.

\section{Summary and outlook}
\label{sec:future}
Statistical representation of a sound model discrepancy term is
certainly challenging. Still, the assumption of \textit{zero} model
discrepancy is a rather extreme position. Almost \textit{any}
reasonable guess is better than nothing in order to avoid false values
for the model parameters and to minimize over-fitting.

The importance of model discrepancies is neatly summarized in the
famous quote of George E. P.  Box: ‘Essentially, all models are wrong,
but some are useful’~\cite{box1987empirical}, with the additional
comment in Ref.~\cite{brynjarsdottir}: 'But a model that is wrong can
only be useful if we acknowledge the fact that it is wrong.'

Fortunately,most of the \textit{ab initio} models of atomic nuclei are
built on methods from EFT, which by construction promises extra
information about the expected impact of the neglected or missing
physics in theoretical predictions. Bayesian inference is a natural
choice for accounting for model discrepancies and prior knowledge,
especially when the priors have a physical basis. Indeed, extracting
the posterior pdf for the model parameters via Bayesian inference
methods makes it possible to abandon the notion of having \textit{a
  single} parameterization of a particular interaction potential and
instead build models based on a continuous pdf of
parameters. Developments along these lines are already taking place in
e.g. density functional theory for atomic
nuclei~\cite{PhysRevLett.122.062502}.

At the moment, most theoretical analyses of atomic nuclei proceed in
the following fashion. Given a potential $V(\vec{\alpha}_{\star})$,
optimized to reproduce some set of calibration data $D$, we setup a
model $M(\vec{\alpha}_{\star},\vec{x})$ to analyze an experimental
result corresponding to the control setting $\vec{x}_i$, i.e. we
evaluate $M(\vec{\alpha}_{\star},\vec{x}_i)$. In a few cases we
propagate uncertainties originating from the measurement errors
present in the data vector $D$, and sometimes we estimate the EFT
truncation error using a series of models at different chiral
orders. This takes a lot of effort. Indeed, \textit{ab initio} nuclear
models are represented by complex computer codes, implemented via
years of dedicated work by several people, and computationally
expensive to evaluate. On top of that, to understand the underlying
nuclear interaction is, arguably, on of the most difficult problems in
all physics. Still, we would like to to answer questions like: how
much should we trust a model prediction? is the model $M$ over-fitted?
why is it not agreeing with observed data, and how do we understand
this discrepancy? 

We should strive to use Bayesian methods for calibrating our models
$M(\vec{\alpha},\vec{x})$ to obtain posterior pdfs
$P(\vec{\alpha}|M,D,I)$ for the parameters. Subsequent evaluations of
an observable $\mathcal{O}_i$, corresponding to setting the model
control variable to $\vec{x}_i$, should be marginalized over the
parameter posterior pdf to produce a posterior predictive pdf
\begin{equation}
  P(\mathcal{O}_{i}|M,\vec{x}_i,D) = \int {\rm d}\vec{\alpha} \,\, P(\mathcal{O}_i|\vec{\alpha},M,\vec{x}_i,D)P(\vec{\alpha}|M,D).
\end{equation}
This quantity will best reflect our state of knowledge, and is quite
meaningful to compare with data. Various marginalizations with respect
to subsets of the parameters can provide better insights into the
qualities of the \textit{ab initio} model. Bayesian inference also
allows us to compare different models via the computation of Bayes
factors~\cite{gelman}, which in turn enables us to address questions
like: which PC in chiral EFT has the strongest support by data? It is
also theoretically straightforward to compute the posterior predictive
pdf averaged over a set of different models $\mathcal{M} =
[M_{1},M_2,\ldots,M_3]$~\cite{hoeting}, each weighted by their
probability of being true, in the finite space spanned by
$\mathcal{M}$, given data $D$.

\subsection{The computational challenge}
There are several challenges connected with the outlook presented
above: working out the theoretical underpinnings of chiral EFT,
specifying prior information, formulating model discrepancy terms, and
performing challenging Markov Chain Monte Carlo
evaluation~\cite{brooks2011handbook} of complicated posterior
pdfs. From a practical point of view, the computational complexity is
the most difficult one. Indeed, evaluating models of medium- and
heavy-mass atomic nuclei typically requires vast high-performance
computing resources. This clearly puts the feasibility of the Bayesian
scenario presented above into question. Without any unforeseen
disruptive computer technologies or dramatic algorithmic advances, it
will be necessary to employ, where possible, fast emulators that
accurately mimic the response of the original \textit{ab initio}
models. This is where we can draw from advances in machine
learning. Possibly useful methods are e.g. Gaussian process regression
and artificial neural networks. Both of these approaches can be
challenging since they introduce hyperparameters that require
additional optimization. Although it can be difficult to assess how
well such methods will work, there exist several examples of useful
surrogate interpolation and extrapolation in nuclear modelling, see
e.g. Refs.~\cite{Ekstr_m_2019, PhysRevC.100.044001,
  PhysRevC.99.054308,
  PhysRevC.100.054326,PhysRevC.98.034318,PhysRevLett.122.062502,McDonnell2015}. Recently,
a new method called eigenvector continuation~\cite{Frame2018} turns
out to be a promising tool for accurate extrapolation and fast
emulation of nuclear properties~\cite{konig2019}. In a recent
paper~\cite{ekstrm2019global}, this method proved capable of
emulating, with a root mean squared error of 1\%, more than one
million solutions of an \textit{ab initio} model for the ground-state
energy and radius of $^{16}$O in one hour on a standard laptop. An
equivalent set of exact \textit{ab initio} coupled-cluster
computations would require 20 years.

\section*{Funding}
This work was supported by the European Research Council (ERC) under
the European Unions Horizon 2020 research and innovation programme
(Grant agreement No. 758027).

\section*{Acknowledgments}
I would like to thank all my collaborators for fruitful discussions
and for sharing their insights during our joint work on the topics
presented here.

\bibliographystyle{frontiersinHLTH&FPHY} 
\bibliography{refs}
\end{document}